\newif\ifsubmission

\ifsubmission
\else
\documentclass{amsart}
\fi

\usepackage[utf8]{inputenc}
\usepackage{amsmath}
\usepackage{amsfonts}
\usepackage{amssymb}
\usepackage{xfrac,color,url}
\usepackage{tikz}
\usetikzlibrary{arrows.meta}

\usepackage[group-separator={,},group-minimum-digits={3}]{siunitx}

\begin{document}

\title[]{Data-driven time-frequency tessellation for signals with oscillatory amplitude envelopes and instantaneous frequency, with application to photoplethysmograhy}

\ifsubmission

\else

\author{Jennifer Laine}

\address{Department of Mathematics, Yale University, New Haven, 06520, CT, USA} 

\author{Hau-Tieng Wu}

\address{Courant Institute of Mathematical Sciences, New York University, New York, 10012, USA} 

\fi

\begin{abstract}
Biomedical signals often comprise multiple non-sinusoidal oscillatory components whose amplitude modulation (AM) and instantaneous frequency (IF) may themselves be governed by additional (second-order) oscillatory dynamics with time-varying amplitude and frequency. 
We introduce a novel time-frequency (TF) analysis framework, {\em Tessellation-based Ensembled Time-Frequency Representation via Integrated Shifting} (TETRIS), designed based on the proposed generalized adaptive non-harmonic model to leverage second-order oscillatory information in this class of signals. We present the model and algorithm using the photoplethysmogram (PPG) as a canonical example, whose cardiac component is known to encode respiratory information in both AM and IF, and demonstrate how respiratory signals can be recovered from PPG.
The central idea of TETRIS is to partition the TF plane along the estimated IF of the cardiac component and to process each partition adaptively to enhance representation quality. This tessellation enables a refined time-frequency representation (TFR), allowing more effective recovery of the respiratory modulation governing the AM of the cardiac component.
We provide theoretical justification for the proposed method and validate its performance on semi-synthetic signals. Finally, we demonstrate that TETRIS enables improved reconstruction of multiple surrogate respiratory signals directly from PPG data. While the model and algorithm are developed with a focus on PPG, the framework is flexible and has potential to be applied to other signals.
\end{abstract}

\ifsubmission
\begin{keywords}
TETRIS \sep SST \sep SAMD \sep PPG \sep RIIV \sep RIAV \sep RIFV
\end{keywords}

\else\fi

\maketitle

\section{Introduction}

Biomedical signals often comprise multiple oscillatory components whose amplitude modulations (AMs) and instantaneous frequencies (IFs) may themselves be driven by additional oscillations. 
Such ``second-order'' oscillations arise in a wide range of scientific fields.
In physiology, electrocardiogram signals exhibit respiratory sinus arrhythmia (RSA), where the instantaneous heart rate (IHR) oscillates with breathing \cite{berntson1997heart}, while neural rhythms (e.g., alpha or theta bands) show amplitude modulation by slower cortical oscillations \cite{canolty2010functional}. 
In acoustics, musical tones display vibrato (periodic frequency modulation) and tremolo (periodic AM) \cite{fletcher2012physics}, and speech signals exhibit jointly varying pitch and amplitude envelopes due to prosody \cite{rosen1992temporal}. 
In mechanical systems, vibration signals from rotating machinery often have amplitudes modulated by periodic impacts, as in faulty bearings or gear meshing \cite{antoni2009cyclostationarity}. 
In engineered systems, amplitude- and frequency-modulated carrier waves in communications \cite{oppenheim1997signals}, as well as micro-Doppler radar returns from moving objects, provide further instances of this structure \cite{chen2006micro}. This is a far from exhaustive list.

In this paper, we explore this class of signals with a focus on the photoplethysmogram (PPG), whose cardiac oscillation is known to encode respiratory information in its AM and IFs via respiratory-induced variations (RIVs), typically categorized as \cite{pimentel2016toward} respiratory-induced intensity variation (RIIV),
respiratory-induced frequency variation (RIFV), and
respiratory-induced amplitude variation (RIAV).
As PPG is widely deployed on mobile and homecare devices, which often come with only a single channel, maximizing information extraction is essential, and respiratory information is a target. 
Numerous methods exploit RIVs to obtain respiratory dynamics; see \cite{charlton2017breathing} for a review, \cite{dehkordi2018extracting,selvakumar2022realtime,iqbal2022photoplethysmography,sultan2023continuous,shuzan2023machine,zhang2025respiratory} for representative approaches for respiratory rate estimate, \cite{cicone2017nonlinear} for instantaneous respiratory rate (IRR) estimate, and recent deep-learning methods \cite{davies2023rapid,shuzan2024ppg2respnet} for respiratory waveform recovery.

Despite this progress, a principled model of PPG that properly describes RIVs is lacking to our knowledge, limiting effective use of the available information. Motivated by wisdom from other fields \cite{rosen1992temporal,berntson1997heart,chen2006micro,antoni2009cyclostationarity,fletcher2012physics},  accumulated experience of prior PPG analyses, and empirical observations of repetitive patterns (see Figure \ref{fig demo real PPG}) in time-frequency representations (TFRs) obtained by time-frequency (TF) methods \cite{lin2016waveshape,cicone2017nonlinear}, we propose a {\em generalized adaptive non-harmonic model} (gANHM) to model PPG with RIIV, RIAV, and RIFV, and explore the nontrivial interaction of RIAV and RIFV. This analysis explains several peculiar behaviors observed in the TFR of PPG (see Figure \ref{fig demo real PPG}). 
Building on this model and recent insights on spectral interference analysis \cite{chand2026spectral}, we introduce a new TF method, {\em Tessellation-based Ensembled Time-frequency Representation via Integrated Shifting} (TETRIS). Applied to PPG, TETRIS exploits the repetitive TFR structure produced by transforms such as the short-time Fourier transform (STFT) and synchrosqueezing transform (SST) \cite{DaLuWu2011} to partition the TF plane along the IHR and its multiples. Each $k$-th region corresponds to the $k$-th harmonic of the cardiac component. The respiratory dynamics, RIAV and RIFV, is encoded in each region. For $k>1$, regions are shifted to the fundamental via the SST reconstruction formula and then averaged. The resulting representation enhances TFR quality by averaging out nonstationary noise impact on different TF regions (see Figure \ref{fig TFR all steps}), which enables stable recovery of oscillatory structure in the cardiac AM. We demonstrate the method on a semi-real signal for illustration of the method, and apply it to reconstruct respiratory signals encoded in PPG via RIIV and RIAV (see Figure \ref{fig reconstructed RESP}).

\section{A Phenomenological Model for PPG}

Before providing a phenomenological model to describe PPG, we start with some physiological facts.

\subsection{Physiological facts}

PPG encodes rich respiratory information via different mechanisms \cite{shelley2006use,shelley2006best,shelley2007photoplethysmography,allen2007photoplethysmography}.
RIIV is the {\em amplitudes change of the PPG peaks} \cite[Section II.A 1)]{pimentel2016toward}. It is believed to be attributed to respiration-driven intrathoracic pressure changes that modulate venous return, vascular compliance, and baseline blood volume, producing a direct current (DC) offset in the signal. Let $t_k$ be the time of the $k$-th systolic peak of the PPG signal $f(t)$. Interpolating the nonuniform samples ${(t_k, f(t_k))}$ (e.g., via cubic splines) yields an upper envelope $E_u(t)$. Bandpass filtering $E_u(t)$ (e.g., 0.1-1 Hz) provides the reconstructed RIIV signal, referred to as the {\em traditional RIIV} (tRIIV) algorithm.
RIAV represents modulation of the cardiac component, or the alternating current (AC), of the PPG, and is defined as the difference in amplitude between the corresponding peak and trough \cite[Section II.A 2)]{pimentel2016toward}. During inspiration, decreased intrathoracic pressure increases venous return to the right heart, while transiently reducing left ventricular stroke volume due to pulmonary blood pooling; expiration reverses this effect. Through the Frank-Starling mechanism, these cyclical preload changes produce oscillations in stroke volume and hence the amplitude variation of the cardiac component. Let $s_k$ denote the trough times. Interpolating ${(s_k, f(s_k))}$ gives a lower envelope $E_l(t)$. We call the bandpass-filtered difference $E_u(t)-E_l(t)$ (e.g., 0.1-1 Hz) the {\em traditional RIAV} (tRIAV) algorithm.
RIFV is modulation of heart rate by respiration \cite[Section II.A 3)]{pimentel2016toward}, known as RSA via cardiopulmonary coupling. Typically, the heart rate increases during inspiration and decreases during expiration.

\subsection{Existing mathematical model and its limitation}
Below, we call a smooth $1$-periodic function $s$ with zero mean and unit $L^2$ norm a {\em wave-shape function} (WSF) \cite{Wu:2013}. A $C^1$ function $T$ is a $\Delta$-trend if $\texttt{supp}\hat{T}\subset [-\Delta/2,\Delta/2]$, where $\Delta\geq 0$ is small. A positive $C^1$ function $f(t)$ is $\epsilon$-slowly varying function if $|f'(t)|\leq \epsilon f(t)$ for some small $\epsilon\geq 0$.

We start with reviewing the existing {\em adaptive non-harmonic model} (ANHM) \cite{Wu:2013,lin2016waveshape}, a phenomenological framework tailored to biomedical signals. We first formulate it for PPG in a way that captures RIIV, and then extend it to incorporate RIAV and RIFV. For small $\epsilon\geq 0$ and $\Delta>0$, we model a PPG signal as
\begin{align}
f_0(t)= A_0(t)s_0(\phi_0(t))+A_1(t)s_1(\phi_1(t))+T_0(t)\label{model time series0}\,,
\end{align}
where $t\in \mathbb{R}$ and, for $l=0,1$:
\ifsubmission
(C1) $A_l(t)$ is positive, $C^1$, and $\epsilon$-slowly varying;
(C2) $\phi_l$ is strictly increasing and $C^2$, with $\phi_l'(t)$ $\epsilon$-slowly varying; 
(C3) $s_l$ is a WSF;
(C4) $\phi_1'(t)-\phi_0'(t)>\Delta$, and $\phi'_0(t)>\Delta/2$;
(C5) $T_0(t)$ is a $\Delta$-trend.
\else
\begin{itemize}
\item (C1) $A_l(t)$ is positive, $C^1$, and $\epsilon$-slowly varying;
\item (C2) $\phi_l$ is strictly increasing and $C^2$, with $\phi_l'(t)$ $\epsilon$-slowly varying; 
\item (C3) $s_l$ is a WSF;
\item (C4) $\phi_1'(t)-\phi_0'(t)>\Delta$, and $\phi'_0(t)>\Delta/2$;
\item (C5) $T_0(t)$ is a $\Delta$-trend.
\end{itemize}
\fi
In this formulation, the low frequency component $A_0(t)s_0(\phi_0(t))$ models RIIV, which leads to the amplitudes change of the PPG peaks used in tRIIV. The high frequency component $A_1(t)s_1(\phi_1(t))$ corresponds the cardiac component, while $T_0(t)$ captures tissue and venous oxygenation information \cite{allen2007photoplethysmography,shelley1993detection}.
We call $A_l(t)$ and $\phi_l(t)$ as the AM and phase functions, respectively, and physiologically, $\phi'_0(t)$ and $\phi'_1(t)$ capture the instantaneous respiratory rate (IRR) and IHR.

\subsection{Proposed phenomenological model}

While \eqref{model time series0} is effective for many PPG analyses and other biomedical signals \cite{lin2016waveshape}, it becomes inadequate when RIAV or RIFV is present in the cardiac component. In practice, we also observe that RIAV modulation varies across different cardiac harmonics. To capture {\em harmonic-dependent} RIAV and RIFV, we expand each WSF in a Fourier series, yielding 
\begin{align}
f_0(t)=\sum_{l=0}^1 \sum_{k=1}^{d_l} A_l(t)\alpha_{l,k}\cos(2\pi k\phi_l(t)+\beta_{l,k})+T_0(t) \,,\label{model time series0-1}
\end{align}
where $\alpha_{l,k}\geq0$ and $\beta_{l,k}\in [0,2\pi)$ are Fourier coefficients of $s_l$ and $d_l\in \mathbb{N}$ is the associated harmonic order. Based on (C3) assumption, $\alpha_{l,1}>0$. 
We then generalize (C1) and (C2) of \eqref{model time series0-1} and model PPG by a random process
\begin{equation}\label{model time series1}
Y(t)= R_0(t)+ C(t) +T_0(t) +\Phi(t)\,,
\end{equation}
where $\Phi$ is a $0$ mean random process with finite variance modeling the inevitable noise, the RIIV becomes
\ifsubmission
$R_0(t):= A_0(t)\sum_{k=1}^{d_0}\alpha_{0,k}\cos(2\pi k\phi_0(t)+\beta_{0,k})$,
\else
$$R_0(t):= A_0(t)\sum_{k=1}^{d_0}\alpha_{0,k}\cos(2\pi k\phi_0(t)+\beta_{0,k}),$$
\fi
and the cardiac component is
\begin{align}
C(t):=\sum_{l=1}^{d_1} A_{1,l}(t)\cos(2\pi l\phi_{1}(t)+\beta_{1,l})\label{model time series1C}\,,
\end{align}
with amplitude and phase
\ifsubmission
\begin{align}\label{model osc Am component}
A_{1,l}(t)=T_l(t)+\sum_{k=1}^{d_0} a_{l,k}(t)\cos(2\pi k\phi_{0}(t)+\beta_{a,k})\ \ \mbox{and}\ \ \phi_{1}(t)=\phi(t)+\frac{b(t)}{2\pi \phi'_0(t)}\sin(2\pi \phi_{0}(t))\,.
\end{align}
\else
\begin{align}\label{model osc Am component}
A_{1,l}(t)&\,=T_l(t)+\sum_{k=1}^{d_0} a_{l,k}(t)\cos(2\pi k\phi_{0}(t)+\beta_{a,k})\\
\phi_{1}(t)&\,=\phi(t)+\frac{b(t)}{2\pi \phi'_0(t)}\sin(2\pi \phi_{0}(t))\,.\nonumber
\end{align}
\fi
Here the following conditions are satisfied  for all $l=1,\ldots,d_1$ and $k=1,\ldots,d_0$:
(G1) $a_{l,k}(t)>0$ and $b(t)>0$ are $\epsilon$-slowly varying $C^1$ functions, $b(t)\leq \delta$ for a small $\delta \in[0,1)$;
(G2) $\phi_0$ and $\phi$ are strictly increasing and $C^2$ with both $\phi_0'(t)$ and $\phi'(t)$ $\epsilon$-slowly varying;
(G3) $\beta_{a,k}\in[0,2\pi)$ are global phases;
(G4) $\phi'_1(t)-d_0\phi'_0(t)>\Delta$;
(G5) $T_l(t)$ are $\Delta$-trend and $A_{1,l}(t)>0$.
In this formulation, $A_{1,l}(t)\cos(2\pi l\phi_{1}(t)+\beta_{1,l})$ is the {\em $l$-th cardiac harmonic}. The component $\sum_{k=1}^{d_0}a_{l,k}(t)\cos(2\pi k\phi_0(t)+\beta_{a,k})$ models RIAV of the $l$-th cardiac harmonic, which we call {\em RIAV-$l$}. The phase $\phi_{1}(t)=\phi(t)+ \frac{b(t)}{2\pi \phi'_0(t)}\cos(2\pi \phi_{0}(t)+\beta_{f})$ models RIFV of the cardiac component, which has IHR 
\ifsubmission
$\phi'_1(t)=\phi'(t)+ b(t)\cos(2\pi \phi_{0}(t)+\beta_{f})$ 
\else
$$\phi'_1(t)=\phi'(t)+ b(t)\cos(2\pi \phi_{0}(t)+\beta_{f})$$ 
\fi
that is modulated by the respiration. For simplicity, the WSF of RIFV is a cosine function. Although this can be generalized to more complex forms, such extensions offer little additional insight and would complicate an already intricate model. We therefore retain this simpler specification. Here, $\delta$ is assumed to be sufficiently small so that $\phi'_1(t)>0$, ensuring physiological plausibility.  
(G4) reflects the typical separation between respiratory and cardiac rates. Except on the extreme scenarios, usually respiratory rate is slower than half of the heart rate. While it can be further generalized, we keep it simple here. Model \eqref{model osc Am component}, together with (C4), (C5), and (G1)-(G5), defines our final PPG model incorporating RIIV, RIAV, and RIFV, which we term the {\em generalized ANHM} (gANHM).
Although $d_0,d_1$ may be infinite in principle, and the correct order needs to be estimated using, for example, linear regression technique \cite{ruiz2022wave}. Based on empirical observation, the magnitude of $A_{1,l}$ ($a_{l,k}$ resp.) decays fast as $l$ ($k$ resp.) increases, $d_0=2$ and $d_1=10$ are usually sufficient to accurately capture the PPG.

To better understand this model, note that we can rewrite 
\ifsubmission
$T_0(t)+R_0(t)= A_{1,0}(t)\cos(2\pi\times 0\times \phi_{1}(t)+\beta_{1,0})$,
\else
$$T_0(t)+R_0(t)= A_{1,0}(t)\cos(2\pi\times 0\times \phi_{1}(t)+\beta_{1,0}),$$
\fi
where $A_{1,0}(t):=T_0(t)+ A_0(t)\sum_{k=1}^{d_0}\alpha_{0,k}\cos(2\pi k\phi_0(t)+\beta_{0,k})$ and $\beta_{1,0}=0$. 
With the trigonometric product-to-sum identity, the RIIF and RIAF in the clean signal in \eqref{model time series1} becomes 
\ifsubmission
\begin{align}
f(t)=&\,\sum_{l=0}^{10} \Big\{ T_l(t)\cos(2\pi l\phi_{1}(t)+\beta_{1,l})+ \frac{1}{2}\sum_{k=1}^{2}a_{l,k}(t) 
\big[\cos(2\pi (l\phi_{1}(t)+k\phi_{0}(t))+\beta_{1,l}+\beta_{a,k})\nonumber\\
&\qquad\qquad + \cos(2\pi (l\phi_{1}(t)-k\phi_{0}(t))+\beta_{1,l}-\beta_{a,k})\big] \Big\}\,, \label{model PPGexpansion2m} 
\end{align}
\else
\begin{align}
f(t)=&\,\sum_{l=0}^{10} \Big\{ T_l(t)\cos(2\pi l\phi_{1}(t)+\beta_{1,l})\nonumber\\
&\qquad+ \frac{1}{2}\sum_{k=1}^{2}a_{l,k}(t) 
\big[\cos(2\pi (l\phi_{1}(t)+k\phi_{0}(t))+\beta_{1,l}+\beta_{a,k})\nonumber\\
&\qquad\qquad + \cos(2\pi (l\phi_{1}(t)-k\phi_{0}(t))+\beta_{1,l}-\beta_{a,k})\big] \Big\}\,, \label{model PPGexpansion2m} 
\end{align}
\fi
which exhibits a clear repetitive structure in the IFs, $l\phi'_{1}(t)+k\phi'_{0}(t)$, symmetric about $l\phi'_1(t)$. Moreover, the amplitudes of oscillations are also symmetric about $l\phi'_1(t)$.
We can continue with RIFV. Denote $\theta_l(t):=\frac{lb(t)}{2\phi'_0(t)}$. When $\delta$ is sufficiently small, for any $l$ we have 
\ifsubmission
$\cos(2\pi l\phi_{1}(t)+\beta_{1,l})
=\cos(2\pi l\phi(t)+\beta_{1,l}+\theta_l(t)\sin(2\pi \phi_{0}(t)))
=\cos(2\pi l\phi(t)+\beta_{1,l})  \cos (\theta_l(t)\sin(2\pi \phi_{0}(t)) )
-\sin(2\pi l\phi(t)+\beta_{1,l}) \sin (\theta_l(t)\sin(2\pi \phi_{0}(t)) )$
\else
\begin{align*}
&\cos(2\pi l\phi_{1}(t)+\beta_{1,l})\\
=\,&\cos(2\pi l\phi(t)+\beta_{1,l}+\theta_l(t)\sin(2\pi \phi_{0}(t)))\\
=\,&\cos(2\pi l\phi(t)+\beta_{1,l})  \cos (\theta_l(t)\sin(2\pi \phi_{0}(t)) )\\
\,&-\sin(2\pi l\phi(t)+\beta_{1,l}) \sin (\theta_l(t)\sin(2\pi \phi_{0}(t)) )
\end{align*}
\fi
 by the trigonometric sum-to-product identity, which becomes 
 \ifsubmission
$\big(1-\theta_l(t)^2\big)\cos(2\pi l\phi(t)+\beta_{1,l})
+\theta_l(t) [\cos(2\pi (l\phi(t)+\phi_{0}(t))+\beta_{1l})-\cos(2\pi (l\phi(t)-\phi_{0}(t))+\beta_{1l}) ]
+\frac{\theta_l(t)^2}{2} [\cos(2\pi (l\phi(t)+2\phi_{0}(t))+\beta_{1l})+\cos(2\pi (l\phi(t)-2\phi_{0}(t))+\beta_{1l})]
+O(\delta^3)$ 
\else
\begin{align*}
&\big(1-\theta_l(t)^2\big)\cos(2\pi l\phi(t)+\beta_{1,l})\\
&+\theta_l(t) [\cos(2\pi (l\phi(t)+\phi_{0}(t))+\beta_{1l})-\cos(2\pi (l\phi(t)-\phi_{0}(t))+\beta_{1l}) ]\\
&+\frac{\theta_l(t)^2}{2} [\cos(2\pi (l\phi(t)+2\phi_{0}(t))+\beta_{1l})+\cos(2\pi (l\phi(t)-2\phi_{0}(t))+\beta_{1l})]
+O(\delta^3)
\end{align*}
\fi
by Taylor expansion of sine and cosine at $0$.   
Similarly, we have
\ifsubmission
$\cos(2\pi (l\phi_{1}(t)+k\phi_0(t))+\beta_{1,l}+\beta_{a,k})
= (1-\theta_l(t)^2 )\cos(2\pi (l\phi(t)+k\phi_0(t))+\beta_{1,l}+\beta_{a,k})
+\theta_l(t)[\cos(2\pi (l\phi(t)+(k+1)\phi_0(t))+\beta_{1,l}+\beta_{a,k})-\cos(2\pi(l\phi(t)+(k-1)\phi_0(t))+\beta_{1,l}+\beta_{a,k})]
+\frac{\theta_l(t)^2}{2}[\cos(2\pi (l\phi(t)+(k+2)\phi_0(t))+\beta_{1,l}+\beta_{a,k})+\cos(2\pi (l\phi(t)+(k-2)\phi_0(t))+\beta_{1,l}+\beta_{a,k})]
+O(\delta^3)$ 
\else
\begin{align*}
&\cos(2\pi (l\phi_{1}(t)+k\phi_0(t))+\beta_{1,l}+\beta_{a,k})\\
= \,& (1-\theta_l(t)^2 )\cos(2\pi (l\phi(t)+k\phi_0(t))+\beta_{1,l}+\beta_{a,k})\\
&+\theta_l(t)[\cos(2\pi (l\phi(t)+(k+1)\phi_0(t))+\beta_{1,l}+\beta_{a,k})\\
&\qquad-\cos(2\pi(l\phi(t)+(k-1)\phi_0(t))+\beta_{1,l}+\beta_{a,k})]\\
&+\frac{\theta_l(t)^2}{2}[\cos(2\pi (l\phi(t)+(k+2)\phi_0(t))+\beta_{1,l}+\beta_{a,k})\\
&\qquad+\cos(2\pi (l\phi(t)+(k-2)\phi_0(t))+\beta_{1,l}+\beta_{a,k})]
+O(\delta^3)
\end{align*}
\fi
and same for $\cos(2\pi (l\phi_{1}(t)-k\phi_0(t))+\beta_{1,l}-\beta_{a,k})$.
In other words, when the amplitude is fixed to $1$, up to $O(\delta^3)$, each cardiac harmonic exhibits a clear repetitive structure in the IFs, $l\phi'_{1}(t)+k\phi'_{0}(t)$ (or side-bands), symmetric about $l\phi'_1(t)$, and the oscillation amplitudes are also symmetric about $l\phi'_1(t)$.
Plugging this into \eqref{model PPGexpansion2m}, a nontrivial interaction of RIAF and RIFV appears and we have
\ifsubmission
 \begin{align}
f(t)= \sum_{l=0}^{10} \sum_{k=-4}^{4} \tilde{a}_{l,k}(t)\cos(2\pi (l\phi(t)+k\phi_0(t))+\gamma_{l,k}) +O(\delta^3) =\sum_{l=0}^{10} \sum_{k=-4}^{4}f_{c,l,k}(t)+O(\delta^3)\,,\label{model PPGexpansion2}
\end{align}
\else
 \begin{align}
f(t)= \,&\sum_{l=0}^{10} \sum_{k=-4}^{4} \tilde{a}_{l,k}(t)\cos(2\pi (l\phi(t)+k\phi_0(t))+\gamma_{l,k}) +O(\delta^3) \nonumber\\
=\,&\sum_{l=0}^{10} \sum_{k=-4}^{4}f_{c,l,k}(t)+O(\delta^3)\,,\label{model PPGexpansion2}
\end{align}
\fi
where 
\ifsubmission
$\tilde{a}_{l,0}(t)= T_l(t) +\theta_l(t)\cos(\beta_{a,1}) +  \theta_l(t)^2\big(\frac{\cos(\beta_{a,2})}{4}- T_l(t)\big)$, $\gamma_{l,0}=\beta_{1,l}$, and by the trigonometric sum-to-product formula, terms of the same IF are aggregated into
$\tilde{a}_{l,1}(t)\cos(2\pi (l\phi(t)+\phi_0(t))+\gamma_{l,1})=
\frac{1-\theta_l(t)^2}{2} a_{l,1}(t)\cos(2\pi (l\phi(t)+\phi_0(t))+\beta_{1,l}+\beta_{a,1})  
+ \theta_l(t) T_l(t)\cos(2\pi (l\phi(t)+\phi_0(t))+\beta_{1,l}) 
- \frac{\theta_l(t)}{2} a_{l,2}(t)\cos(2\pi(l\phi(t)+\phi_0(t))+\beta_{1,l}+\beta_{a,2})
+ \frac{\theta_l(t)^2}{4}\cos(2\pi (l\phi(t)+\phi_0(t))+\beta_{1,l}-\beta_{a,1})$,
$\tilde{a}_{l,-1}(t)\cos(2\pi (l\phi(t)+\phi_0(t))+\gamma_{l,-1})=
\frac{1-\theta_l(t)^2}{2} a_{l,1}(t)\cos(2\pi (l\phi(t)-\phi_0(t))+\beta_{1,l}-\beta_{a,1})  
- \theta_l(t) T_l(t)\cos(2\pi (l\phi(t)-\phi_0(t))+\beta_{1,l}) 
+ \frac{\theta_l(t)}{2} a_{l,2}(t)\cos(2\pi(l\phi(t)-\phi_0(t))+\beta_{1,l}-\beta_{a,2})
+ \frac{\theta_l(t)^2}{4}\cos(2\pi (l\phi(t)-\phi_0(t))+\beta_{1,l}+\beta_{a,1})$,
$\tilde{a}_{l,2}(t)\cos(2\pi (l\phi(t)+2\phi_0(t))+\gamma_{l,2})=
\frac{1-\theta_l(t)^2}{2} a_{l,2}(t) \cos(2\pi (l\phi(t)+2\phi_0(t))+\beta_{1,l}+\beta_{a,2}) 
+ \frac{\theta_l(t)}{2} a_{l,1}(t)\cos(2\pi (l\phi(t)+2\phi_0(t))+\beta_{1,l}+\beta_{a,1})$,
$\tilde{a}_{l,-2}(t)\cos(2\pi (l\phi(t)+2\phi_0(t))+\gamma_{l,-2})=
\frac{1-\theta_l(t)^2}{2}  a_{l,2}(t) \cos(2\pi (l\phi(t)-2\phi_0(t))+\beta_{1,l}-\beta_{a,2}) 
- \frac{\theta_l(t)}{2} a_{l,1}(t)\cos(2\pi (l\phi(t)-2\phi_0(t))+\beta_{1,l}-\beta_{a,1})$,
$\tilde{a}_{l,3}(t)\cos(2\pi (l\phi(t)+3\phi_0(t))+\gamma_{l,3})=\frac{\theta_l(t)}{2} a_{l,2}(t)\cos(2\pi (l\phi(t)+3\phi_0(t))+\beta_{1,l}+\beta_{a,2})
+\frac{\theta_l(t)^2}{4} a_{l,1}(t)\cos(2\pi (l\phi(t)+3\phi_0(t))+\beta_{1,l}+\beta_{a,1})$,
$\tilde{a}_{l,-3}(t)\cos(2\pi (l\phi(t)+3\phi_0(t))+\gamma_{l,-3})=-\frac{\theta_l(t)}{2} a_{l,2}(t)\cos(2\pi (l\phi(t)-3\phi_0(t))+\beta_{1,l}-\beta_{a,2})
+\frac{\theta_l(t)^2}{4} a_{l,1}(t)\cos(2\pi (l\phi(t)-3\phi_0(t))+\beta_{1,l}-\beta_{a,1})$,
$\tilde{a}_{l,4}(t)\cos(2\pi (l\phi(t)+3\phi_0(t))+\gamma_{l,4})=\frac{\theta_l(t)^2}{4} a_{l,2}(t)\cos(2\pi (l\phi(t)+4\phi_0(t))+\beta_{1,l}+\beta_{a,2})$, 
and
$\tilde{a}_{l,-4}(t)\cos(2\pi (l\phi(t)+4\phi_0(t))+\gamma_{l,-4})=\frac{\theta_l(t)^2}{4} a_{l,2}(t)\cos(2\pi (l\phi(t)-4\phi_0(t))+\beta_{1,l}-\beta_{a,2})$.
\else
\[
\tilde{a}_{l,0}(t)= T_l(t) +\theta_l(t)\cos(\beta_{a,1}) +  \theta_l(t)^2\big(\frac{\cos(\beta_{a,2})}{4}- T_l(t)\big),
\] 
$\gamma_{l,0}=\beta_{1,l}$, and by the trigonometric sum-to-product formula, terms of the same IF are aggregated into
\allowdisplaybreaks
\begin{align*}
&\tilde{a}_{l,1}(t)\cos(2\pi (l\phi(t)+\phi_0(t))+\gamma_{l,1})\\
=&\,\frac{1-\theta_l(t)^2}{2} a_{l,1}(t)\cos(2\pi (l\phi(t)+\phi_0(t))+\beta_{1,l}+\beta_{a,1})  \\
&+ \theta_l(t) T_l(t)\cos(2\pi (l\phi(t)+\phi_0(t))+\beta_{1,l}) \\
&- \frac{\theta_l(t)}{2} a_{l,2}(t)\cos(2\pi(l\phi(t)+\phi_0(t))+\beta_{1,l}+\beta_{a,2})\\
&+ \frac{\theta_l(t)^2}{4}\cos(2\pi (l\phi(t)+\phi_0(t))+\beta_{1,l}-\beta_{a,1})\,,\\
&\tilde{a}_{l,-1}(t)\cos(2\pi (l\phi(t)+\phi_0(t))+\gamma_{l,-1})\\
=&\,\frac{1-\theta_l(t)^2}{2} a_{l,1}(t)\cos(2\pi (l\phi(t)-\phi_0(t))+\beta_{1,l}-\beta_{a,1})  \\
&- \theta_l(t) T_l(t)\cos(2\pi (l\phi(t)-\phi_0(t))+\beta_{1,l})\\ 
&+ \frac{\theta_l(t)}{2} a_{l,2}(t)\cos(2\pi(l\phi(t)-\phi_0(t))+\beta_{1,l}-\beta_{a,2})\\
&+ \frac{\theta_l(t)^2}{4}\cos(2\pi (l\phi(t)-\phi_0(t))+\beta_{1,l}+\beta_{a,1})\,,\\
&\tilde{a}_{l,2}(t)\cos(2\pi (l\phi(t)+2\phi_0(t))+\gamma_{l,2})\\
=\,&\frac{1-\theta_l(t)^2}{2} a_{l,2}(t) \cos(2\pi (l\phi(t)+2\phi_0(t))+\beta_{1,l}+\beta_{a,2}) \\
&+ \frac{\theta_l(t)}{2} a_{l,1}(t)\cos(2\pi (l\phi(t)+2\phi_0(t))+\beta_{1,l}+\beta_{a,1})\,,\\
&\tilde{a}_{l,-2}(t)\cos(2\pi (l\phi(t)+2\phi_0(t))+\gamma_{l,-2})\\
=\,&\frac{1-\theta_l(t)^2}{2}  a_{l,2}(t) \cos(2\pi (l\phi(t)-2\phi_0(t))+\beta_{1,l}-\beta_{a,2}) \\
&- \frac{\theta_l(t)}{2} a_{l,1}(t)\cos(2\pi (l\phi(t)-2\phi_0(t))+\beta_{1,l}-\beta_{a,1})\,,\\
&\tilde{a}_{l,3}(t)\cos(2\pi (l\phi(t)+3\phi_0(t))+\gamma_{l,3})\\
=\,&\frac{\theta_l(t)}{2} a_{l,2}(t)\cos(2\pi (l\phi(t)+3\phi_0(t))+\beta_{1,l}+\beta_{a,2})\\
&+\frac{\theta_l(t)^2}{4} a_{l,1}(t)\cos(2\pi (l\phi(t)+3\phi_0(t))+\beta_{1,l}+\beta_{a,1})\,,\\
&\tilde{a}_{l,-3}(t)\cos(2\pi (l\phi(t)+3\phi_0(t))+\gamma_{l,-3})\\
=\,&-\frac{\theta_l(t)}{2} a_{l,2}(t)\cos(2\pi (l\phi(t)-3\phi_0(t))+\beta_{1,l}-\beta_{a,2})\\
&+\frac{\theta_l(t)^2}{4} a_{l,1}(t)\cos(2\pi (l\phi(t)-3\phi_0(t))+\beta_{1,l}-\beta_{a,1})\,,\\
&\tilde{a}_{l,4}(t)\cos(2\pi (l\phi(t)+3\phi_0(t))+\gamma_{l,4})\\
=\,&\frac{\theta_l(t)^2}{4} a_{l,2}(t)\cos(2\pi (l\phi(t)+4\phi_0(t))+\beta_{1,l}+\beta_{a,2})\,,\\
&\tilde{a}_{l,-4}(t)\cos(2\pi (l\phi(t)+4\phi_0(t))+\gamma_{l,-4})\\
=\,&\frac{\theta_l(t)^2}{4} a_{l,2}(t)\cos(2\pi (l\phi(t)-4\phi_0(t))+\beta_{1,l}-\beta_{a,2})\,.
\end{align*}
\fi
In this expression, $\tilde{a}_{l,k}$ are for $C^1$ $\epsilon$ slowly varying and  $\gamma_{l,k}\in [0,2\pi)$ are associated phases. This expansion exhibits a clear repetitive structure $l\phi_{1}(t)+k\phi_{0}(t)$, symmetric about $l\phi_1(t)$, but the amplitudes of $f_{c,l,1}(t)$ and $f_{c,l,-1}(t)$ are in general different, which comes from the nontrivial interaction between RIAV and RIFV. We call $\sum_{k=1}^4f_{c,l,k}$ the {\em surrogate respiratory signal by RIAV and RIFV} of the $l$-th cardiac harmonic and $f_{c,l,k}$ its $k$-th harmonic.

\section{Time-frequency tessellation and proposed algorithm}

Below, when we apply TF analysis, including STFT and SST, the window function is assumed to be $h_L$, a Gaussian window spanning over $L>0$ seconds with the half time support 6.

\subsection{Peculiar repetitive pattern in TFR of PPG signal}

\begin{figure}[hbt!]
    \centering
    \ifsubmission
\includegraphics[trim=0 0 0 0, width=\columnwidth]{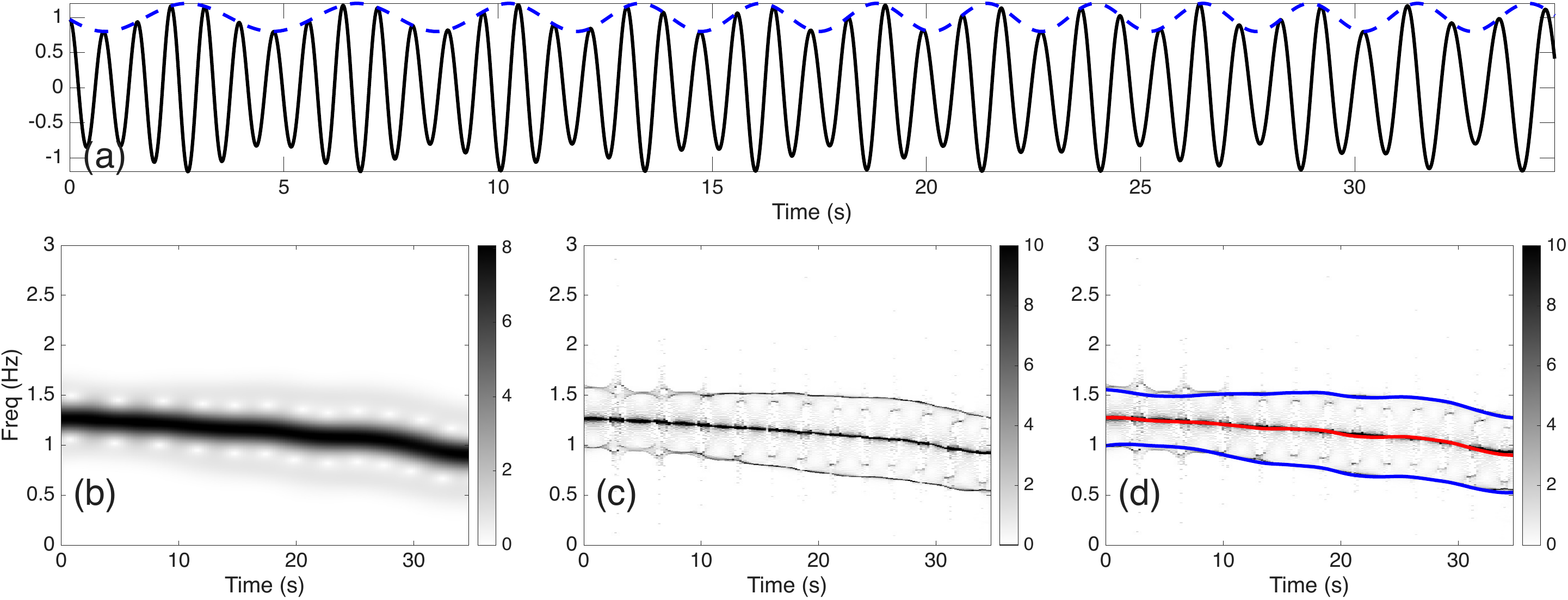}
\else
\makebox[\columnwidth][c]{
\includegraphics[trim=0 0 0 0, width=1.3\columnwidth]{Figure1.pdf}
}
\fi
    \caption{Example of a signal satisfying the gANHM. (a) The signal, with its amplitude modulation (AM) overlaid as a blue dashed curve. (b) The TFR obtained via STFT. (c) The TFR obtained via SST. (d) Same as (c), with $\phi_1'(t)$ superimposed in red and $\phi_1'(t)\pm\phi_2'(t)$ in blue.}
    \label{fig clean signal non-AHM}
\end{figure}

Before introducing the algorithm, we consider a toy example that illustrates how RIAV behaves. 
Consider a toy example with oscillatory behavior of AM:
\ifsubmission
$g(t) = (1+0.2\cos(2\pi\varphi_2(t)))\cos(2\pi\varphi_1(t))$,
\else
\[
g(t) = (1+0.2\cos(2\pi\varphi_2(t)))\cos(2\pi\varphi_1(t)),
\]
\fi
where $\varphi'_1(t)>\varphi_2'(t)>0$, $\varphi_i$ are $\epsilon$-slowly varying, and the WSF is $\cos(2\pi t)$. This function is captured by the gANHM. See Figure \ref{fig clean signal non-AHM} for its behavior.
By the trigonometric product-to-sum identity, $g(t) = \cos(2\pi\varphi_1(t))+0.1\cos(2\pi(\varphi_1(t)-\varphi_2(t)))+0.1\cos(2\pi(\varphi_1(t)+\varphi_2(t)))$. It is well known that when $\epsilon$ and the spectral support of window $h_L$ are sufficiently small, the STFT of $g$, defined as $V_g^{(h_L)}(t,\xi):=\int_{-\infty}^\infty g(x)h_L(x-t)e^{i2\pi \xi(x-t)}dx$, where $(t,\xi)\in \mathbb{R}\times \mathbb{R}^+$, satisfies \cite{DaLuWu2011}
\ifsubmission
$V_g^{(h_L)}(t,\xi)
= \frac{1}{2}\hat{h}_L(\xi-\varphi'_1(t))e^{i2\pi\varphi_1(t)}+\frac{1}{20}[\hat{h}_L(\xi-\varphi'_1(t)-\varphi'_2(t))e^{i2\pi(\varphi_1(t)+\varphi_2(t))}+
\hat{h}_L(\xi-\varphi'_1(t)+\varphi'_2(t))e^{i2\pi(\varphi_1(t)-\varphi_2(t))}]$,
\else
\begin{align*}
V_g^{(h_L)}(t,\xi)
= \,& \frac{1}{2}\hat{h}_L(\xi-\varphi'_1(t))e^{i2\pi\varphi_1(t)}+\frac{1}{20}[\hat{h}_L(\xi-\varphi'_1(t)-\varphi'_2(t))e^{i2\pi(\varphi_1(t)+\varphi_2(t))}\\
&+\hat{h}_L(\xi-\varphi'_1(t)+\varphi'_2(t))e^{i2\pi(\varphi_1(t)-\varphi_2(t))}],
\end{align*}
\fi
with an error of order $\epsilon$. This calculation explains the ``side curves'' above and below $\phi'_1(t)$ in the TFR shown in Figure \ref{fig clean signal non-AHM}.

\begin{figure}[hbt!]
    \centering
    \ifsubmission
    \begin{tikzpicture}
\node (img) {\includegraphics[trim=0 0 0 0, width=\columnwidth]{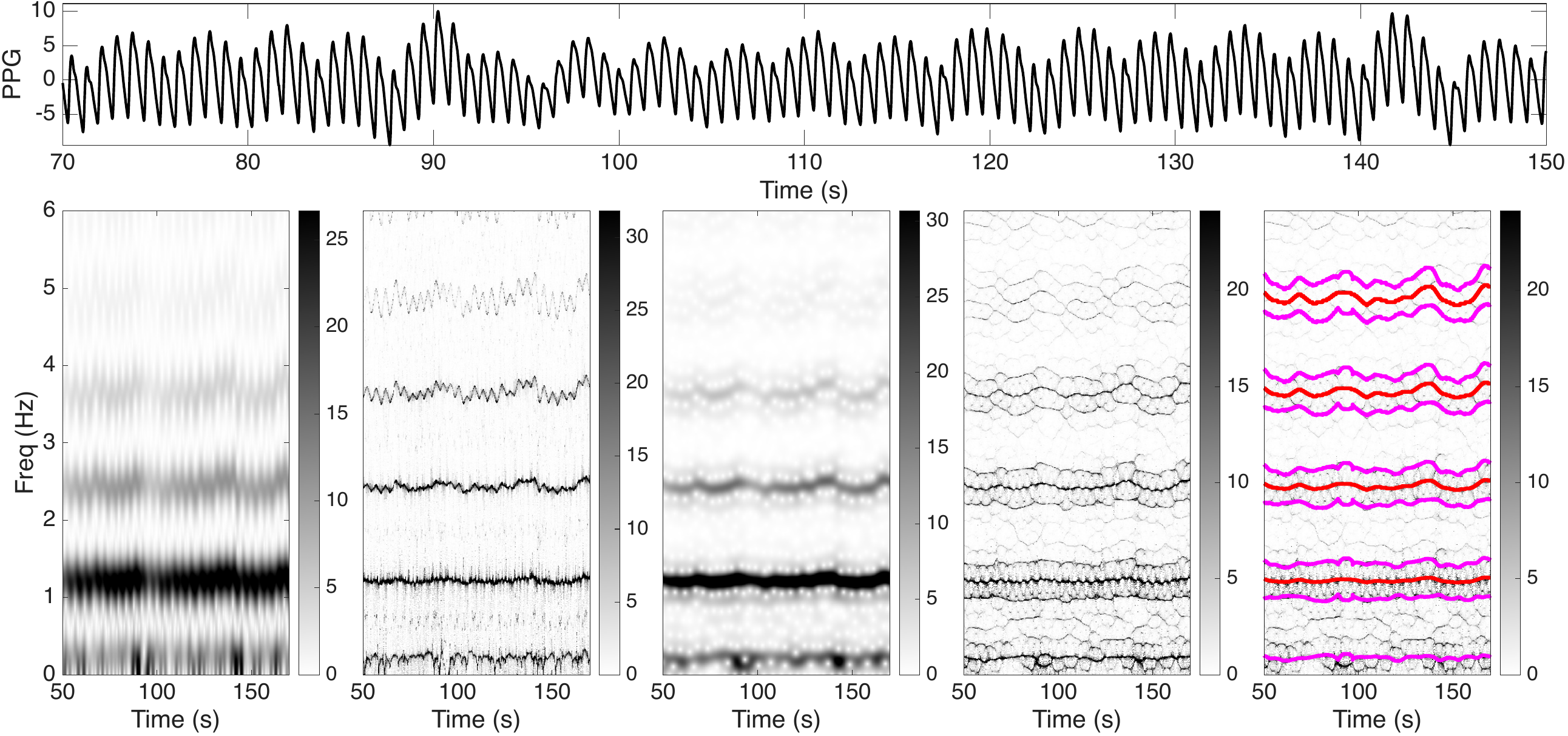}};
\draw[->, red, very thick] (-0.3,-2.8) -- (0.0,-2.5);
\draw[->, red, very thick] (2.8,-2.8) -- (3.1,-2.5);
\draw[-{Stealth}, blue, very thick] (-3.4,-2.7) -- (-3.1,-2.4);
\draw[-{Stealth}, blue, very thick] (-6.5,-2.7) -- (-6.2,-2.4);
\draw[-{Triangle}, magenta, very thick] (2.7,0.4) -- (3.1,0.8);
\end{tikzpicture}
\else
\makebox[\columnwidth][c]{
\begin{tikzpicture}
\node (img) {\includegraphics[trim=0 0 0 0, width=1.3\columnwidth]{71_74DEMOTFR.pdf}};
\draw[->, red, very thick] (-0.3,-2.8) -- (0.0,-2.5);
\draw[->, red, very thick] (2.8,-2.8) -- (3.1,-2.5);
\draw[-{Stealth}, blue, very thick] (-3.4,-2.7) -- (-3.1,-2.4);
\draw[-{Stealth}, blue, very thick] (-6.5,-2.7) -- (-6.2,-2.4);
\draw[-{Triangle}, magenta, very thick] (2.7,0.4) -- (3.1,0.8);
\end{tikzpicture}
}
\fi
    \caption{A real PPG signal is shown in the top panel (scaled by $1/1000$). Bottom panels, from left to right: STFT with a 10 s window, SST with a 10 s window, STFT with a 28 s window, SST with a 28 s window, and SST with a 28 s window with the IHR and its harmonics overlaid in red, and the IRR together with $l$-th cardiac harmonics shifted by $\pm$ IRR overlaid in magenta. The IRR is confirmed with the simultaneously recorded abdominal movement.}
    \label{fig demo real PPG}
\end{figure}

The situation is more complicated in practice. 
Under the gANHM, the RIAV causes each cardiac harmonic to ``split'' into $2d_0+1$ components  \eqref{model PPGexpansion2}, which is further complicated by RIFV \eqref{model PPGexpansion2}, where each term in  over $l$ has slowly varying AM and IF. See Figure \ref{fig demo real PPG} for a representative PPG signal with RIIV, RIAV, and RIFV (also \cite[Figure 9]{lin2016waveshape} and \cite[Figure 1]{cicone2017nonlinear}). The characteristic wavy modulation within each cardiac cycle is clearly visible and corresponds to tRIIV. 
Following the notation in \eqref{model PPGexpansion2}, the STFT and SST with $h_{10}$ leads to spurious vertical structures (indicated by blue arrows) near $\phi'(t)$. These arise from spectral interference between $f_{c,l,0}(t)$ and $f_{c,l,\pm1}(t)$ due to the too short window (see \cite{chand2026spectral} for a theoretical treatment). 
When the window is $h_{28}$, $f_{c,l,0}(t)$ and $f_{c,l,\pm1}(t)$ become visually separable, and the dominant ridges align with $l\phi'(t)\pm \phi_0'(t)$, as shown in the lower-right panel. Note that on the TFR, the side-band of $\phi'(t)+ \phi_0'(t)$ is weaker than that of $\phi'(t)- \phi_0'(t)$, which is caused by the interaction between RIAV and RIFV. There is a regular ``bubble''-like structure between $\phi'(t)$ and $\phi'(t) \pm \phi_0'(t)$ coming from the spectral interference, and this interference becomes more irregular and pronounced when examining higher-order cardiac harmonics due to fast IF changes, as indicated by the magenta arrow.

\subsection{Tessellation-based Ensembled TFR via Integrated Shifting (TETRIS)}

Due to the non-constant IHR and IRR, higher-order cardiac harmonics exhibit increasingly rapid frequency variation, leading to pronounced spectral interference and low quality TFR. As a result, direct reconstruction of the oscillatory AM of these higher-order cardiac harmonics becomes challenging.
To effectively utilize this information, and motivated by the TFR in Figure \ref{fig demo real PPG} (and the semi-real simulation in Figure \ref{fig clean signal semi-real PPG} below), a natural question arises: can we exploit the symmetric structure associated with each cardiac harmonic and the regularly spaced pattern of their IFs?
We answer this question in the affirmative and propose the following tessellation-based algorithm.

Consider the gANHM and its expression \eqref{model PPGexpansion2}. Suppose the window function $h$ satisfies $\texttt{supp}(\hat{h})\subset [-\Delta/2,\Delta/2]$.  {\em Ideally}, the associated TFR determined by STFT can be expressed as
\begin{align} 
V^{(h)}_f\,&(t,\xi)
= \frac{1}{2}\sum_{l=0}^{10}\sum_{k=-4}^4 \tilde{a}_{l,k}(t) \hat{h}(\xi-(l\phi'(t)+k\phi'_0(t)))e^{i2\pi(l\phi(t)+k\phi_0(t))+\gamma_{l,k}} 
 \label{model PPG ideal STFT}
\end{align}
for $(t,\xi)\in \mathbb{R}\times \mathbb{R}^+$ up to error of $\epsilon$ order. 
Unfortunately, this approximation is valid only in regions where $l\phi'_{1}(t)\pm k\phi'_{0}(t)$ varies sufficiently slowly. See Figure \ref{fig demo real PPG} for an illustration.

To resolve this issue, note that if the ideal representation in \eqref{model PPG ideal STFT} holds, the TF domain can be partitioned so that each region exhibits a repetitive structure. Specifically, we introduce the data-driven partition
\[
\mathbb{R}\times \mathbb{R}^+=\cup_{j=1}^{d_1} D_j \cup D_{d_1+1}\,,
\]
where for 
\ifsubmission
$D_j:=\{(t,\xi)|\, t\in \mathbb{R},\, \xi\in ((j-1)\phi'_1(t), j\phi'_1(t)]\}$ for $j=1,\ldots, d_1$ and $D_{d_1+1}:=\{(t,\xi)|\, t\in \mathbb{R},\, \xi\in (d_l\phi'_1(t),\infty)\}$.
\else
\[
D_j:=\{(t,\xi)|\, t\in \mathbb{R},\, \xi\in ((j-1)\phi'_1(t), j\phi'_1(t)]\}
\]
 for $j=1,\ldots, d_1$ and 
 \[
 D_{d_1+1}:=\{(t,\xi)|\, t\in \mathbb{R},\, \xi\in (d_l\phi'_1(t),\infty)\}.
 \]
\fi
Clearly, $D_l\cap D_j=\emptyset$ for $l\neq j$, and each $D_j$ can be viewed as a vertical shift of $D_{j-1}$ by $\phi'_1(t)$ at each time instant, for $j=2,\ldots,d_1$. We call this construction a {\em TF tessellation}. 
For $l=,1\ldots,d_l$, denote the boundary of $D_l$ by $\partial D_l:=U_l\cup L_l$, where $U_l=\{(t,l\phi'_1(t)\}$ and $L_l=\{(t, (l-1)\phi'_1(t)\}$. Under this tessellation, within each $D_l$, the RIAV-$l$ and RIFV jointly manifest as IFs $l\phi'_1(t)-k\phi'_0(t)$ and the RIAV-$(l-1)$ and RIFV jointly manifest as $(l-1)\phi'_1(t)+k\phi'_0(t)$, $k=1,\ldots,d_0$, which exhibit a consistent pattern relative to $\partial D_l$.  

This observation inspired us to consider the following {\em phase shifting scheme}. By the reconstruction formula of SST \cite[(13)(14)]{wu2025uncertainty} we can accurately estimate $2\pi\phi_{1}(t)+\beta_{1,1}$. Evaluate a complex-valued function
\[
\tilde{f}_{\downarrow1}(t):= f(t)e^{-i2\pi (\phi_{1}(t)+\beta_{1,1})}\,,
\]
where $T_0(t)$ and $R_0(t)$ became $T_0(t)e^{-i2\pi (\phi_{1}(t)+\beta_{1,1})}$ and $A_0(t)\sum_{k=1}^{d_0}\alpha_{0,k}(e^{i (2\pi (k\phi_0(t)-\phi_1(t))+\beta_{0,k})-\beta_{1,1}}+e^{-i (2\pi (k\phi_0(t)+\phi_1(t))+\beta_{0,k}+\beta_{1,1})})$ in $\tilde{f}_{\downarrow1}(t)$. These terms have ``negative'' IF, and their appearance in the TFR determined by STFT is negligible. The first cardiac component is ``shifted down'' so that terms related to $l=1$ in \eqref{model PPGexpansion2} becomes
\begin{align}
\frac{1}{4}\sum_{k=1}^{d_0}a_{1,k}(t)e^{i (2\pi k\phi_{0}(t)+\beta_{a,k})}+o.w.\label{model PPGexpansion3}\,,
\end{align}
where we only keep the oscillatory components with positive frequencies and the $o.w.$ term includes $\frac{1}{2}T_1(t)$ and all components that have negative IF, including $ \frac{1}{2}T_1(t) e^{-i2\pi 2(\phi_{1}(t)+\beta_{1,1})} +\frac{1}{4}\sum_{k=1}^{d_0}a_{1,k}(t) \big[e^{i (-2\pi k\phi_{0}(t)-\beta_{a,k})}+ e^{-i (2\pi (2\phi_1(t)+k\phi_{0}(t))+2\beta_{1,1}+\beta_{a,k})} + e^{-i (2\pi (2\phi_1(t)-k\phi_{0}(t))+2\beta_{1,1}-\beta_{a,k})}\big]$. For other $l>1$, the cardiac harmonic's behavior is ``shifted'' to behave like that of the $(l-1)$th cardiac harmonic of $f$.
As a result, the TFRs of $\tilde{f}_{\downarrow1}(t)$ and $f$ are similar over $D_1\cup D_2$, since their oscillatory components share the same IHR$\pm$IRR caused by RIAV and RIFV. 

We can iterate by setting
\[
\tilde{f}_{\downarrow k}(t):= \tilde{f}_{\downarrow k-1}(t) e^{-i2\pi (\phi_{1}(t)+\beta_{1,k}-\beta_{1,k-1})}\,,
\]
where $\phi_{1}(t)+\beta_{1,k}-\beta_{1,k-1}$ is estimated from $\tilde{f}_{\downarrow k-1}(t)$ via the reconstruction formula. Similarly, the TFRs of $\tilde{f}_{\downarrow k}(t)$ and $\tilde{f}_{\downarrow k-1}(t)$ are comparable over $D_1\cup D_2$, although the representation deteriorates due to the weaker higher-order cardiac harmonic and the inevitable noise.

Assume $R_f(t,\xi)$ is a TFR determined by either STFT, SST, or others, define the {\em Tessellation-based Ensembled Time-frequency Representation via Integrated Shifting} (TETRIS) by
\begin{equation}
T(t,\xi):=\left(\frac{1}{Q+1}\sum_{k=0}^{Q} w_k \left|R_{\tilde{f}_{\downarrow k}}(t,\xi)\right|^p\right)^{1/p}\,,\label{TETRIS formula}
\end{equation}
where $Q\leq d_1-m$ if we want to obtain higher quality TFR over $\cup_{l=1}^mD_l$, $m\geq 1$, $w_k>0$ is the chosen weight, and $p\geq 1$ is the chosen norm. In practice, since cardiac harmonics higher than 5 are usually weak, for PPG analysis, we suggest to set $Q=5$.
While it is potential to optimize weights and norm, we continue the discussion with $w_k=1$ and $p=1$ below, and leave the optimization to future empirical work. See Figures \ref{fig clean signal semi-real PPG} and \ref{fig TFR all steps} for an illustration of how TETRIS works in a semi-real and real PPG signals.

\subsection{Surrogate respiratory signal reconstruction algorithm}

With TETRIS, we propose the following algorithm to reconstruct surrogate respiratory signal from PPG. 

\begin{itemize}
\item (Step 0: Pre-processing)
Continuous PPG signals are uniformly sampled at  $f_s = 50$Hz for $T>60$ second long and high-pass filtered with a fourth-order Butterworth IIR filter (cutoff 0.1 Hz). Denote the filtered PPG as $\mathbf{f}_{\downarrow0} \in \mathbb{R}^N$. Cardiac cycles are identified via peak detection. Let $t_k$ denote the sample index of the $k$-th cardiac cycle. The IHR is estimated by cubic spline interpolation of $\{(t_k,1/(t_k-t_{k-1}))\}_{k=2}^{n_c}$, where $n_c\in \mathbb{N}$ is the number of cardiac cycles, denoted as $\tilde{\phi}'_1\in \mathbb{R}^N$. 

\item (Step 1: Construct TETRIS)
Denote the TFR of $\mathbf{f}_{\downarrow0}$ by the SST with window $h_L$ \cite{DaLuWu2011} as $S_{L,\mathbf{f}_{\downarrow0}}\in \mathbb{C}^{M\times N}$, where $M=\lfloor N/2\rfloor$ is the number of positive-frequency bins. 
Using $\tilde{\phi}'_1$ as a reference, apply ridge extraction and reconstruction formula \cite{wu2025uncertainty} to  $S_{10,\mathbf{f}}$ to obtain an estimate of $\phi_1(t)+\frac{\beta_{1,1}}{2\pi}$, denoted as $\tilde{\psi}_{1,0}\in \mathbb{R}^N$.
In addition, compute $S_{90,\mathbf{f}_{\downarrow0}}$. 

Then, iterate the following steps for $k=1,\ldots,Q$. 
Compute $\mathbf{f}_{\downarrow k}:=\mathbf{f}_{\downarrow k-1}e^{-i2\pi\tilde{\psi}_{1,k-1}}\in \mathbb{C}^N$, where the product of two vectors is evaluated entrywisely. Calculate $S_{10,\mathbf{f}_{\downarrow k}}$, and apply the ridge extraction and reconstruction formula to $S_{10,\mathbf{f}_{\downarrow k}}$ to obtain an estimate of $\phi_1(t)+\frac{\beta_{1,k}-\beta_{1,k-1}}{2\pi}$, denoted as $\tilde{\psi}_{1,k}\in \mathbb{R}^N$. Similarly, compute $S_{90,\mathbf{f}_{\downarrow k}}$.
Denote the TETRIS as $T_{90}:=\frac{1}{Q+1}\sum_{k=0}^{Q}|S_{90,\mathbf{f}_{\downarrow k}}|\in \mathbb{R}_+^{M\times N}$. The TETRIS $T_{10}\in \mathbb{R}_+^{M\times N}$ is constructed for sanity check.

\item (Step 2: Estimate RIIV and RIAV)
Apply ridge extraction over $0.1-0.5$ Hz band to $T_{90}$ to obtain an estimate of the IRR, denoted $\tilde{\phi}'_0\in \mathbb{R}^N$.
For each $l=0,1,\ldots,Q$, use $\tilde{\phi}'_0$ as the reference and apply ridge extraction to $S_{90,\mathbf{f}_{\downarrow l}}$, followed by reconstruction algorithm to estimate the AM and phase of the first harmonic associated with RIIV (when $l=0$) or the first harmonic of the surrogate respiratory signal by RIAV and RIFV of the $l$ th cardiac harmonic (when $l>0$), denoted as $\tilde{a}_{R,l}\in \mathbb{R}^N$ and $\tilde{\phi}_{R,l}\in \mathbb{R}^N$. Then, apply shape-adaptive mode decomposition (SAMD) \cite{colominas2021decomposing} with $\tilde{a}_{R,l}$ and $\tilde{\phi}_{R,l}$ to reconstruct the RIIV (when $l=0$) or the surrogate respiratory signal by RIAV and RIFV of the $l$-th cardiac harmonic (when $l>0$), denoted as $\tilde{R}_l\in \mathbb{R}^N$. In SAMD, the harmonic and the polynomial orders are both set to $2$.

\end{itemize}

Note that $\tilde{\phi}'_1$ is an estimate of RIFV. The parameters are chosen according to physiological considerations. The window lengths $L=10$ and $L=90$ in SST correspond approximately to 10 cardiac cycles and 15 respiratory cycles, respectively. Since RIIV and surrogate respiratory signal by RIAV and RIFV are relatively weak compared with the cardiac component, we purposely demand more cycles to enhance the stability of the algorithm. The spectral band $0.1-0.5$ Hz in Step 2 is selected based on the normal adult respiratory rate. In SAMD, the polynomial order should remain low; order 2 is empirically found to be both stable and accurate. None of these parameters are optimized in this work; in practice, optimal tuning could be achieved via data-driven learning, which we leave for future empirical study.
While in this paper TETRIS and reconstruction algorithm are motivated by and developed for PPG, we believe it is applicable to other similar biomedical signals. We leave this generalization to future work.

\subsection{Theoretical consideration}

Consider the model \eqref{model time series1}. We claim that TETRIS can help reduce the noise impact and lead to a higher quality TFR. Assume $\phi_1(t)=\xi_0t$ and $\Phi$ is stationary Gaussian random process with the spectral density 
\ifsubmission
$p(\xi)\asymp (1+|\xi|)^\rho$, 
\else
\[
p(\xi)\asymp (1+|\xi|)^\rho,
\] 
\fi
where $\rho<5$. Denote the shifted noise as $\Phi_{\downarrow 1}(t):= \Phi(t)e^{-i2\pi\xi_0t}$. For a pair $(t,\xi)$, it is well known that $V^{(h)}_\Phi(t,\xi)$ and $V^{(h)}_{\Phi_{\downarrow 1}}(t,\xi)$ are both centered complex Gaussian random variables. Denote $h_{t,\xi}(x):=h(x-t)e^{-i2\pi \xi (x-t)}$. Their covariance is 
\ifsubmission
$\mathbb{E}[V^{(h)}_\Phi(t,\xi)\overline{V^{(h)}_{\Phi_{\downarrow 1}}(t,\xi)}]=\int \mathcal{F}(h_{t,\xi})(\eta)\overline{\mathcal{F}(h_{t,\xi} e^{-i2\pi\xi_0\cdot})(\eta)} p(\eta)d\eta=\int \mathcal{F}(h_{t,\xi})(\eta)\overline{\mathcal{F}(h_{t,\xi})(\eta+\xi_0)}p(\eta) d\eta$, 
\else
\begin{align*}
\mathbb{E}[V^{(h)}_\Phi(t,\xi)\overline{V^{(h)}_{\Phi_{\downarrow 1}}(t,\xi)}]=\,&\int \mathcal{F}(h_{t,\xi})(\eta)\overline{\mathcal{F}(h_{t,\xi} e^{-i2\pi\xi_0\cdot})(\eta)} p(\eta)d\eta\\
=\,&\int \mathcal{F}(h_{t,\xi})(\eta)\overline{\mathcal{F}(h_{t,\xi})(\eta+\xi_0)}p(\eta) d\eta,
\end{align*} 
\fi
where $\mathcal{F}$ is Fourier transform. When $\texttt{supp}(\hat{h})$ is sufficiently small compared with $\xi_0$, the supports of $\mathcal{F}(h_{t,\xi})(\eta)$ and ${\mathcal{F}(h_{t,\xi})(\eta+\xi_0)}$ do not overlap and hence 
\ifsubmission
$\mathbb{E}[V^{(h)}_\Phi(t,\xi)\overline{V^{(h)}_{\Phi_{\downarrow 1}}(t,\xi)}]=0$. 
\else
\[
\mathbb{E}[V^{(h)}_\Phi(t,\xi)\overline{V^{(h)}_{\Phi_{\downarrow 1}}(t,\xi)}]=0.
\] 
\fi
Similarly, the pseudocovariance 
\ifsubmission
$\mathbb{E}[V^{(h)}_\Phi(t,\xi){V^{(h)}_{\Phi_{\downarrow 1}}(t,\xi)}]=0$. 
\else
\[
\mathbb{E}[V^{(h)}_\Phi(t,\xi){V^{(h)}_{\Phi_{\downarrow 1}}(t,\xi)}]=0.
\] 
\fi
Therefore, $V^{(h)}_{\Phi\downarrow j}(t,\xi)$, $j=0,1\ldots,d_1$, are independent, and similarly for the associated SST coefficients. On the other hand, $V^{(h)}_{f\downarrow j}(t,\xi)$, $j=0,1,\ldots,d_1$, share similar TFR structure. Putting them all together, TETRIS in \eqref{TETRIS formula} leads to a cleaner TFR since the noise component is canceled out by law of large number. 
In practice, noise is seldom stationary Gaussian, phases are typically nonlinear, and phase estimation errors accumulate throughout the procedure. 
Nevertheless, if the phase estimation error is sufficiently controlled, the above argument remains valid. For instance, let $\Phi_{\downarrow 1}(t)$ be $\Phi(t)e^{-i2\pi\tilde{\xi}_0t}$, where $\tilde{\xi}_0$ is an estimated frequency. If $|\tilde{\xi}_0-\xi_0|$ is small enough so that the supports of $\mathcal{F}(h_{t,\xi})(\eta)$ and ${\mathcal{F}(h_{t,\xi})(\eta+\tilde{\xi}_0)}$ do not overlap, the same independence and hence the noise cancellation hold, although the TFR of the clean signal might degrade depending on the error magnitude. 
The general phase case can be addressed via a standard Taylor expansion due to the slowly varying assumption, combined with handling weakly dependent noise in a law of large number framework. We omit details here.
Although it is known that, under mild moment and dependence conditions, the STFT of general nonstationary, non-Gaussian noise converges to a complex Gaussian field \cite{wu2025uncertainty}, the above law-of-large-numbers argument does not directly apply due to the induced dependence structure. A systematic study of TETRIS under such general models is deferred to future work.

\section{Results}
\begin{figure}[hbt!]
    \centering
    \ifsubmission
    \begin{tikzpicture}
\node (img) {\includegraphics[trim=0 0 0 0, width=\columnwidth]{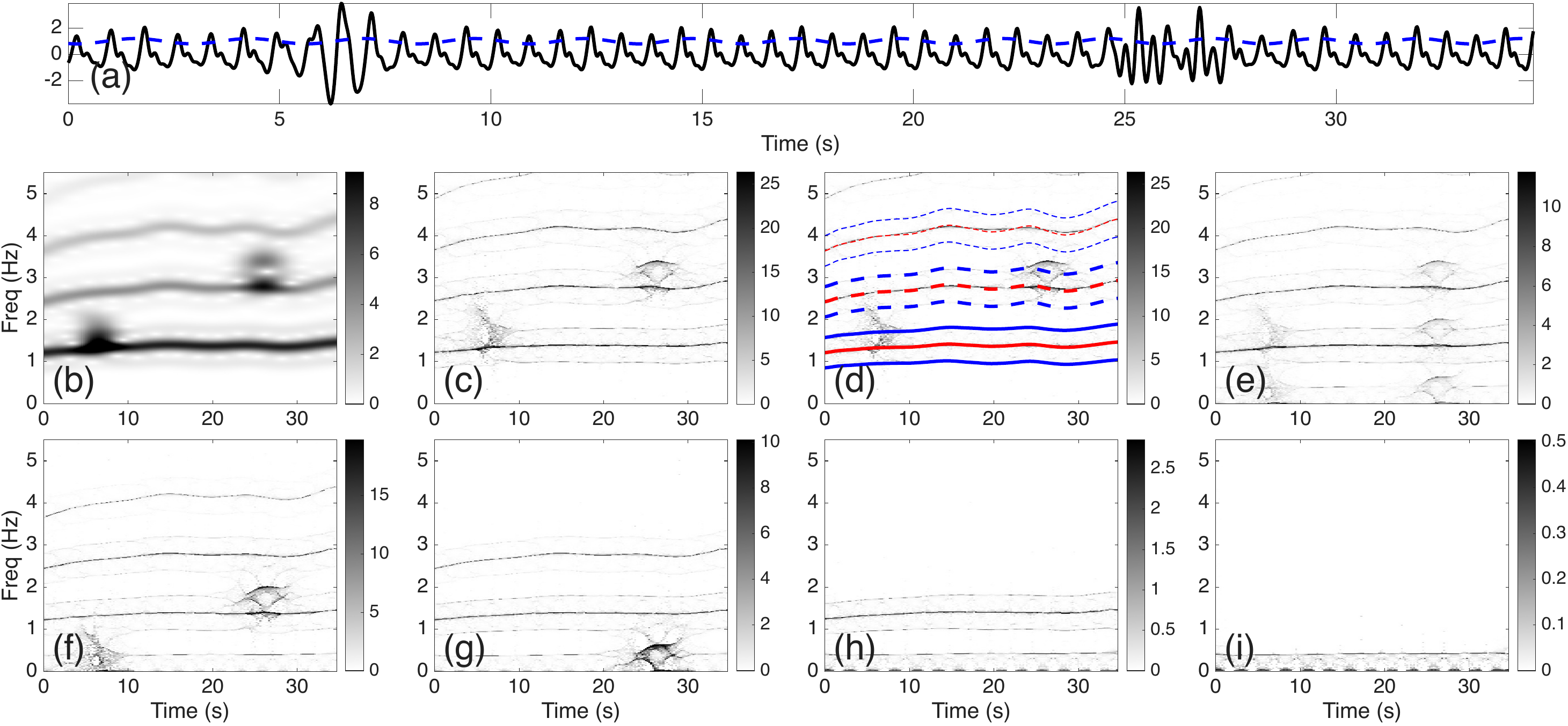}};
\draw[->, red, thick] (-1.3,-0.3) -- (-1,0.0);
\draw[->, red, thick] (-1.3,0.3) -- (-1,0.6);
\draw[->, red, thick] (-1.3,0.9) -- (-1,1.2);
\draw[->, blue, thick] (1.2,1) -- (1.5,1.3);
\draw[->, blue, thick] (2,1) -- (2.3,1.3);
\draw[->, magenta, thick] (-7.5,-3) -- (-7.2,-2.7);
\draw[->, magenta, thick] (-5.9,-2.9) -- (-5.6,-2.6);
\draw[->, magenta, thick] (6.3,-.1) -- (6.7,0.2);
\draw[->, magenta, thick] (4.8,-.2) -- (5.1,0.1);
\end{tikzpicture}
\else
\makebox[\columnwidth][c]{
\begin{tikzpicture}
\node (img) {\includegraphics[trim=0 0 0 0, width=1.3\columnwidth]{Figure2.pdf}};
\draw[->, red, thick] (-1.3,-0.3) -- (-1,0.0);
\draw[->, red, thick] (-1.3,0.3) -- (-1,0.6);
\draw[->, red, thick] (-1.3,0.9) -- (-1,1.2);
\draw[->, blue, thick] (1.2,1) -- (1.5,1.3);
\draw[->, blue, thick] (2,1) -- (2.3,1.3);
\draw[->, magenta, thick] (-7.5,-3) -- (-7.2,-2.7);
\draw[->, magenta, thick] (-5.9,-2.9) -- (-5.6,-2.6);
\draw[->, magenta, thick] (6.3,-.1) -- (6.7,0.2);
\draw[->, magenta, thick] (4.8,-.2) -- (5.1,0.1);
\end{tikzpicture}
}
\fi
    \caption{Example of a semi-real PPG with RIAV. Here, (a) is the signal, with the RIAV superimposed as the blue dashed curve; (b) is the TFR determined by STFT; (c) is the TFR determined by SST, where the superimposed red arrows indicates the curve $k\phi'_1(t)-\phi'_0(t)$, for $k=1,2,3$; (d) replicates (c), with $\phi'_1(t)$ and $\phi'_1(t)\pm \phi'_0(t)$ superimposed as solid red and blue curves, $2\phi'_1(t)$ and $2\phi'_1(t)\pm \phi'_0(t)$ as dashed curves, and $3\phi'_1(t)$ and $3\phi'_1(t)\pm \phi'_0(t)$ as thin dashed curves. Blue arrows mark $3\phi'_1(t)$ and its associated ridge; (e) is the TETRIS; (f) TFR of $\mathbf{f}_{\downarrow 1}$ determined by SST; (g) is the TFR of $\mathbf{f}_{\downarrow 2}$ determined by SST;  (h) is the TFR of $\mathbf{f}_{\downarrow 3}$ determined by SST; (i) is the TFR of $\mathbf{f}_{\downarrow 4}$ determined by SST.}
    \label{fig clean signal semi-real PPG}
\end{figure}

The Matlab code for TETRIS and the RIIV and RIAV reconstruction can be found in \url{https://github.com/hautiengwu2/TETRIS.git}.
We consider a semi-real PPG signal $f(t)= (1+0.2\cos(2\pi\phi_0(t)))(\cos(2\pi \phi_1(t))+0.5\cos(2\pi 2\phi_1(t)+1)+0.3\cos(2\pi 3\phi_1(t)+1.3)+0.1\cos(2\pi 4\phi_1(t)+0.3)$, where $\phi_1'(t)$ and $\phi_0'(t)$ are both $\epsilon$-slowly varying and simulate IHR at approximately 1.3 Hz and IRR at approximately 0.3Hz respectively. The global phases of the harmonics are adapted from a real PPG signal. There is no RIIV nor RIFV, and the RIAV is $1+0.2\cos(2\pi\phi_0(t))$ with the WSF $\cos(2\pi t)$. We then add random noise $\Phi_1(t)+\Phi_2(t)$, where $\Phi_1(t)$ is localized around 6 s with spectral support near 2 Hz, and $\Phi_2(t)$ is localized around 25 s with spectral support near 3 Hz. Denote the resulting signal by $Y(t):=f(t)+\Phi_1(t)+\Phi_2(t)$. Its TFRs, obtained by STFT and SST, TETRIS, as well as those of the shifted signals $\mathbf{Y}_{\downarrow k}$, $k=1,2,3$, are shown in Figure \ref{fig clean signal semi-real PPG}. From panels (b) and (c), the RIAV associated with higher cardiac harmonics is visibly blurred (red arrows). This degradation arises because higher harmonics have larger instantaneous frequencies, reducing the ability of STFT and SST to resolve $k\phi_1'(t)$ from $k\phi_1'(t)\pm \phi_0'(t)$. In addition, the ridges of higher-order harmonics deviate more from the true instantaneous frequency, particularly over 8-12 s (blue arrows).
This limitation is alleviated once we ``shift down'' $Y(t)$ by $e^{-i2\pi k\phi_1(t)}$. Panels (f)-(i) show results for $k=1,\ldots,4$. After shifting down, the RIAV of each harmonic becomes more clearly separated from the trend; for instance, panel (f) shows a marked improvement for the first harmonic, and similar improvements are observed in (g)-(i). Note that the added noise play the role of showing the frequency-shifting effect: in (f), the spectrum of $\Phi_1(t)$ shifts to $D_1$ and vanishes in (g), while the spectrum of $\Phi_2(t)$ shifts to $D_2$ in (f), to $D_1$ in (g), and disappears in (h). Visually, TETRIS provides higher quality TFR, particularly over $D_1\cup D_2$.

\begin{figure}[hbt!]
    \centering
    \ifsubmission
    \begin{tikzpicture}
\node (img) {\includegraphics[trim=0 0 0 0, width=\columnwidth]{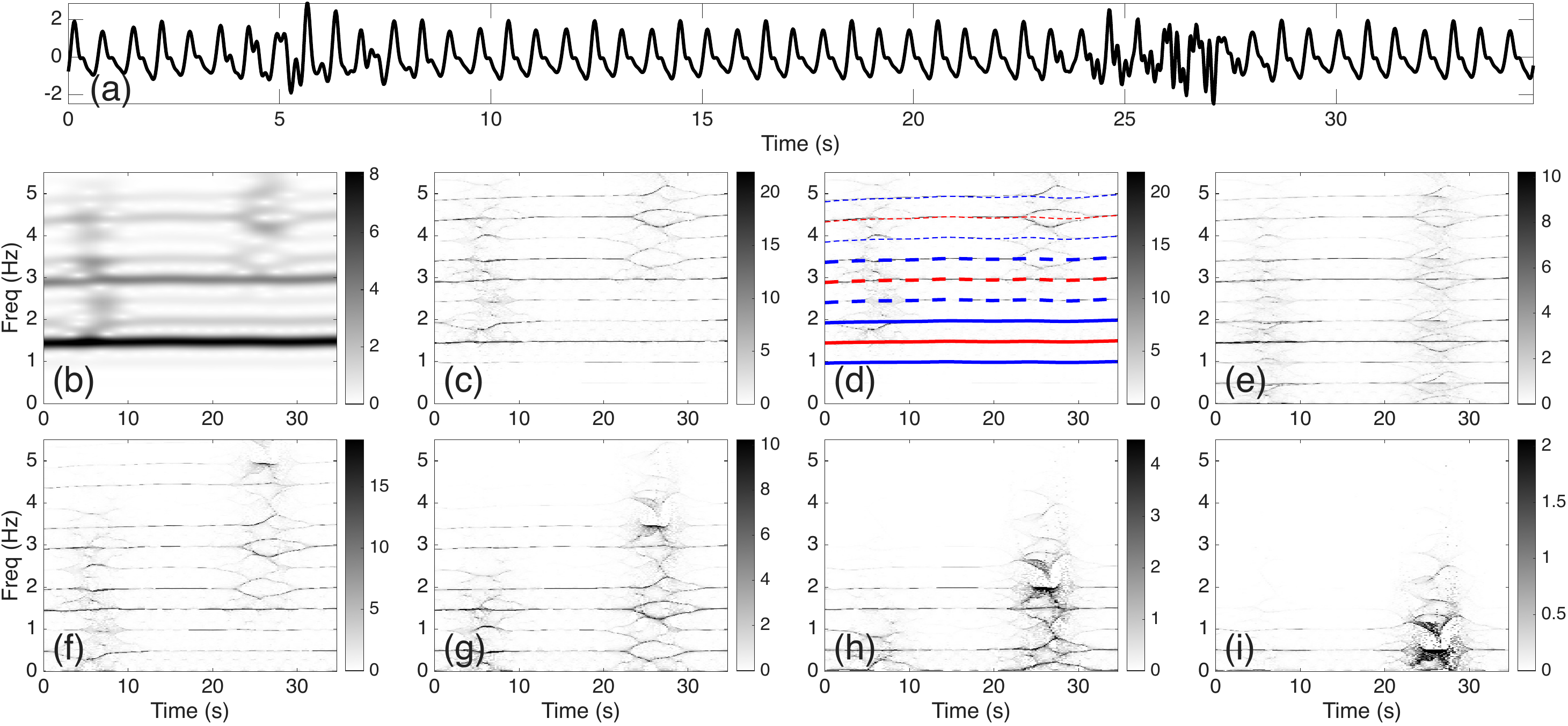}};
\draw[->, red, thick] (-1.7,-0.3) -- (-1.4,0.0);
\draw[->, red, thick] (-1.7,0.1) -- (-1.4,0.4);
\draw[->, red, thick] (-6,-0.3) -- (-5.7,0.0);
\draw[->, red, thick] (-6,0.1) -- (-5.7,0.4);
\draw[->, blue, thick] (6.5,0.1) -- (6.8,0.4);
\draw[->, blue, thick] (-5.8,-2.7) -- (-5.5,-2.4);
\end{tikzpicture}
\else
\makebox[\columnwidth][c]{
\begin{tikzpicture}
\node (img) {\includegraphics[trim=0 0 0 0, width=1.3\columnwidth]{Figure3.pdf}};
\draw[->, red, thick] (-1.7,-0.3) -- (-1.4,0.0);
\draw[->, red, thick] (-1.7,0.1) -- (-1.4,0.4);
\draw[->, red, thick] (-6,-0.3) -- (-5.7,0.0);
\draw[->, red, thick] (-6,0.1) -- (-5.7,0.4);
\draw[->, blue, thick] (6.5,0.1) -- (6.8,0.4);
\draw[->, blue, thick] (-5.8,-2.7) -- (-5.5,-2.4);
\end{tikzpicture}
}
\fi
    \caption{Example of a semi-real PPG with RIAV and RIFV. Here, (a) is the signal; (b) is the TFR determined by STFT with red arrows indicating $\phi'_1(t)\pm\phi'_0(t)$; (c) is the TFR determined by SST with red arrows indicating $\phi'_1(t)\pm\phi'_0(t)$; (d) replicates (c), with $\phi'_1(t)$ and $\phi'_1(t)\pm \phi'_0(t)$ superimposed as solid red and blue curves, $2\phi'_1(t)$ and $2\phi'_1(t)\pm \phi'_0(t)$ as dashed curves, and $3\phi'_1(t)$ and $3\phi'_1(t)\pm \phi'_0(t)$ as thin dashed curves. Blue arrows mark $3\phi'_1(t)$ and its associated ridge; (e) is the TETRIS; (f) TFR of $\mathbf{f}_{\downarrow 1}$ determined by SST; (g) is the TFR of $\mathbf{f}_{\downarrow 2}$ determined by SST;  (h) is the TFR of $\mathbf{f}_{\downarrow 3}$ determined by SST; (i) is the TFR of $\mathbf{f}_{\downarrow 4}$ determined by SST.}
    \label{fig clean signal semi-real PPG2}
\end{figure}

Next, consider a more realistic semi-real PPG signal with RIAV and RIFV, 
\ifsubmission
$f(t)= (1+0.2\cos(2\pi\phi_0(t)+0.5))\times(\cos(2\pi \phi_1(t))+0.5\cos(2\pi 2\phi_1(t)+1)+0.2\cos(2\pi 3\phi_1(t)+1.3)+0.05\cos(2\pi 4\phi_1(t)+0.3)$, 
\else
\begin{align*}
f(t)= &\,(1+0.2\cos(2\pi\phi_0(t)+0.5))\times(\cos(2\pi \phi_1(t))+0.5\cos(2\pi 2\phi_1(t)+1)\\
&+0.2\cos(2\pi 3\phi_1(t)+1.3)+0.05\cos(2\pi 4\phi_1(t)+0.3),
\end{align*} 
\fi
where $\phi_1(t)=\phi(t)+\frac{0.1}{2\pi\phi'_0(t)}\sin(2\pi\phi_0(t))$, $\phi'(t)$ and $\phi_0'(t)$ are both $\epsilon$-slowly varying and simulate IHR at approximately 1.2 Hz and IRR at approximately 0.4Hz respectively. The RIFV is $1+0.1\cos(2\pi\phi_0(t))$ with the WSF $\cos(2\pi t)$. We then add random noise $\Phi_1(t)+\Phi_2(t)$, where $\Phi_1(t)$ is localized around 6 s with spectral support near 3 Hz, and $\Phi_2(t)$ is localized around 25 s with spectral support near 5 Hz. Denote the resulting signal by $Y(t):=f(t)+\Phi_1(t)+\Phi_2(t)$. See Figure \ref{fig clean signal semi-real PPG2} for an analysis result of one realization of $Y(t)$. The same findings from Figure \ref{fig clean signal semi-real PPG} apply here, except that the TFR intensity along $\phi'(t)+\phi_0'(t)$ is stronger than along $\phi'(t)-\phi_0'(t)$, due to the nontrivial interaction between RIAV and RIFV. Visually, TETRIS yields higher TFR quality than the others, particularly in the regions indicated by blue arrows.

\begin{figure}[hbt!]
    \centering
    \ifsubmission
\includegraphics[trim=0 0 0 0, width=\columnwidth]{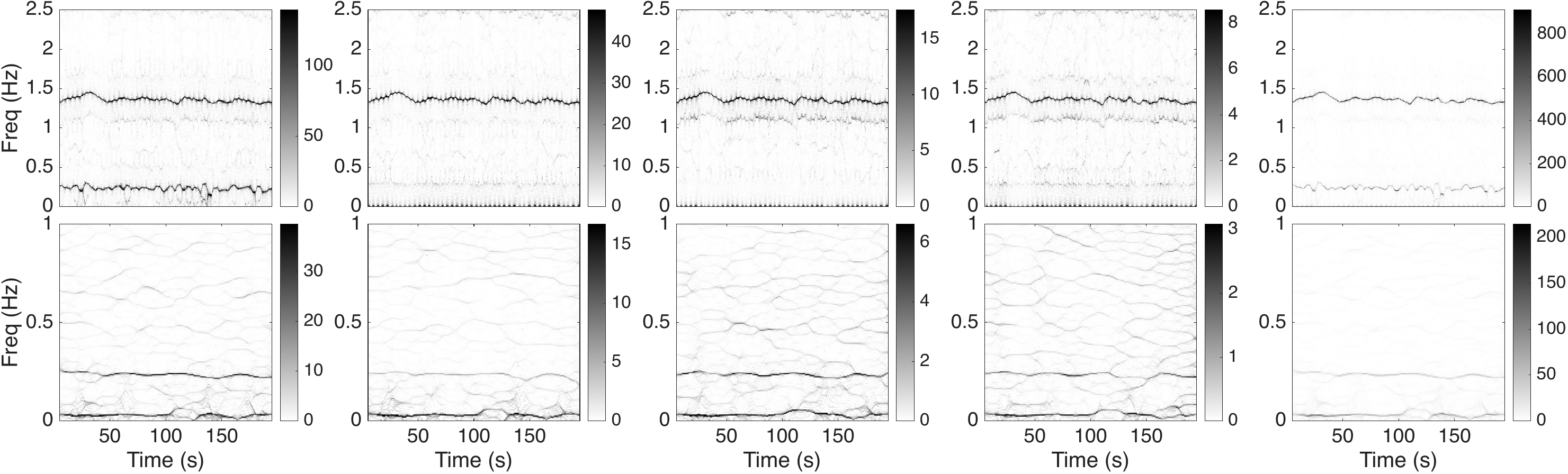}
\else
\makebox[\columnwidth][c]{%
\includegraphics[trim=0 0 0 0, width=1.4\columnwidth]{71_5TFR_ALLsteps.pdf}
}
\fi
    \caption{Top, from left to right, $S_{10,\mathbf{f}_{\downarrow0}}$, $S_{10,\mathbf{f}_{\downarrow1}}$, $S_{10,\mathbf{f}_{\downarrow2}}$, $S_{10,\mathbf{f}_{\downarrow3}}$, and $T_{10}$ of the PPG signal shown in Figure \ref{fig reconstructed RESP}  (scaled by $1/1000$). Bottom, from left to right, $S_{90,\mathbf{f}_{\downarrow1}}$, $S_{90,\mathbf{f}_{\downarrow2}}$, $S_{90,\mathbf{f}_{\downarrow3}}$, $S_{90,\mathbf{f}_{\downarrow4}}$, and $T_{90}$ of the same PPG signal.}
    \label{fig TFR all steps}
\end{figure}

\begin{figure}[hbt!]
    \centering
    \ifsubmission
\includegraphics[trim=0 0 0 0, width=.75\columnwidth]{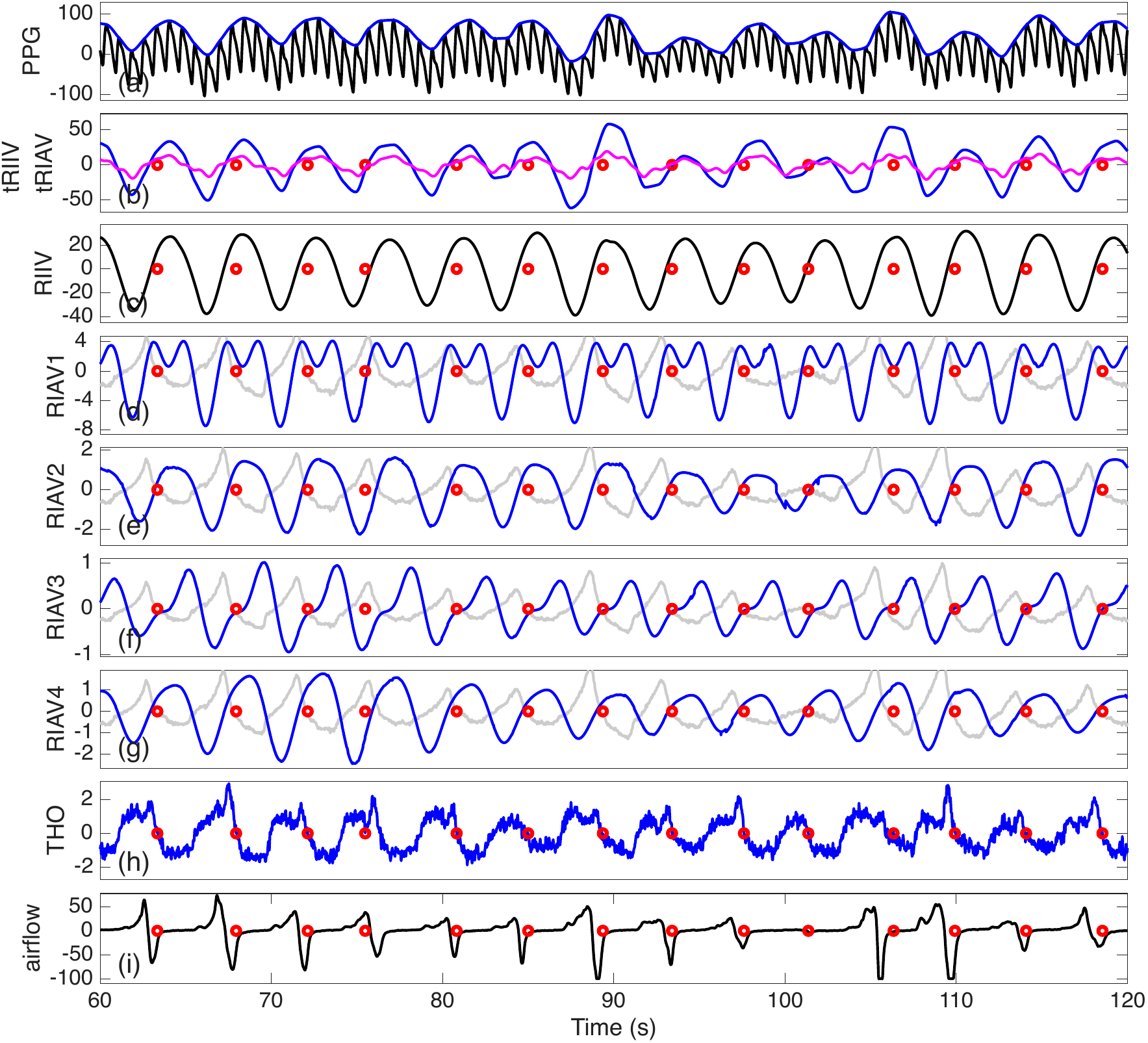}
\else
\makebox[\columnwidth][c]{%
\includegraphics[trim=0 0 0 0, width=1.2\columnwidth]{71_5ExtractedRESP.pdf}
}
\fi
    \caption{Various reconstructed respiratory waveform from PPT. In (b), the blue curve is the traditional RIIV (tRIIV) and the magenta curve is the traditional RIAV (tRIAV). ABD is superimposed as the gray curve in (d)-(g). The red circles are detected expiration starting times from THO.}
    \label{fig reconstructed RESP}
\end{figure}

Next, we consider real PPG signals recorded using the TipTraQ device \cite{chen2025validation} together with simultaneously acquired polysomnography (PSG) from an observational study conducted at Taipei Veterans General Hospital, Taiwan, approved by the Medical Ethics Committee (IRB No. 2023-04-003A). See \cite{chen2025validation} for details. TipTraQ PPG signals were sampled at 50 Hz. Synchronization between TipTraQ and PSG recordings was performed using instantaneous heart rates derived from PSG ECG and TipTraQ PPG. The airflow, thoracic (THO), and abdominal (ABD) respiratory signals from PSG are used as ground truth for comparison. 
Figure \ref{fig TFR all steps} shows TETRIS and intermediate TFRs for a PPG segment under normal breathing. Visually, TETRIS exhibits markedly improved quality compared with the corresponding SST representation, with substantially reduced background noise and enhanced contrast (see colorbar). The operator $T_{90}$ enables a high-quality estimate of the IRR, which in turn supports more accurate reconstruction of RIIV and RIAV. In this segment, a clear low-frequency component around 0.3 Hz is visible in $S_{10,\mathbf{f}_{\downarrow0}}$, corresponding to RIIV. The component below and above IHR, particularly, $S_{10,\mathbf{f}_{\downarrow2}}$ and $S_{10,\mathbf{f}_{\downarrow3}}$, are RIAV encoded in the AM of cardiac harmonics. These structures are not present in $T_{10}$. Also note that in top left panel, the TFR intensity along $\phi'(t)-\phi_0'(t)$ is stronger than along $\phi'(t)+\phi_0'(t)$, which comes from the nontrivial interaction between RIAV and RIFV.
Figure \ref{fig reconstructed RESP} presents the reconstructed RIIV and RIAV signals alongside the original PPG. While tRIIV and tRIAV also reflect respiratory cycles, their morphological fidelity is lower despite bandpass filtering (0.1-1 Hz). For practical respiratory rate estimation algorithms \cite{dehkordi2018extracting,selvakumar2022realtime,iqbal2022photoplethysmography,sultan2023continuous,shuzan2023machine,zhang2025respiratory}, the proposed method has the potential to enhance performance by leveraging higher-quality morphological features. Around 100 s, a brief flow limitation appears in the airflow signal, while oscillations persist in the thoracic and abdominal signals as well as in tRIIV, tRIAV, RIIV, and RIAV. This discrepancy is consistent with the physiological origins of RIIV and RIAV.

\bibliographystyle{plain}
\bibliography{referenceQ}

\end{document}